\documentclass[sn-mathphys,Numbered]{sn-jnl}
\usepackage{graphicx}%
\usepackage{multirow}%
\usepackage{amsmath,amssymb,amsfonts}%
\usepackage{amsthm}%
\usepackage{mathrsfs}%
\usepackage[title]{appendix}%
\usepackage{xcolor}%
\usepackage{textcomp}%
\usepackage{manyfoot}%
\usepackage{booktabs}%
\usepackage{algorithm}%
\usepackage{algorithmicx}%
\usepackage{algpseudocode}%
\usepackage{listings}%
\usepackage{graphicx}
\usepackage{subcaption}

\theoremstyle{thmstyleone}%

\theoremstyle{thmstyletwo}%

\theoremstyle{thmstylethree}%

\begin{document}
	
	\title[Article Title]{Angular Location of the $n^{th}$ Einstein Ring at Large $n$.}
	
	\author*[1,2]{\fnm{Spandan} \sur{Minwalla}}
	\email{spandan.minwalla@gmail.com}
	
	\affil*[1]{\orgname{Aditya Birla World Academy}, \orgaddress{\street{Vastu Shilp, JD Road Annexe}, \city{Mumbai}, \postcode{400007}, \state{Maharashtra}, \country{India}}}
	\affil*[2]{\orgname{École Polytechnique}, \orgaddress{Rte de Saclay, 91120 Palaiseau, France}}

	\abstract{We perform a matched asymptotic expansion to find an analytic formula for the trajectory of a light ray in a Schwarzschild metric, in a power series expansion in the deviation of the impact parameter from its critical value. 
		We present results valid to second sub leading order in this expansion. We use these results to find an analytic expansion for the angular location of the $n^{th}$ Einstein Ring (at large $n$) resulting from a star that lies directly behind a black hole but not necessarily far from it. The small parameter for this expansion is  $e^{-\pi (2n+1) }$: our formulae are accurate to third order in this parameter.}  
	
	\maketitle
	\section{Introduction}

	The remarkable images of the Event Horizon Telescope  \cite{EventHorizonTelescope:2019pgp} highlight the importance of the study of trajectories of light in black hole backgrounds. One obtains valuable insights from the study of the simplest black hole, namely the spherically symmetric Schwarzschild black hole.
	
\begin{figure}
		\centering
		\includegraphics[width=0.7\linewidth]{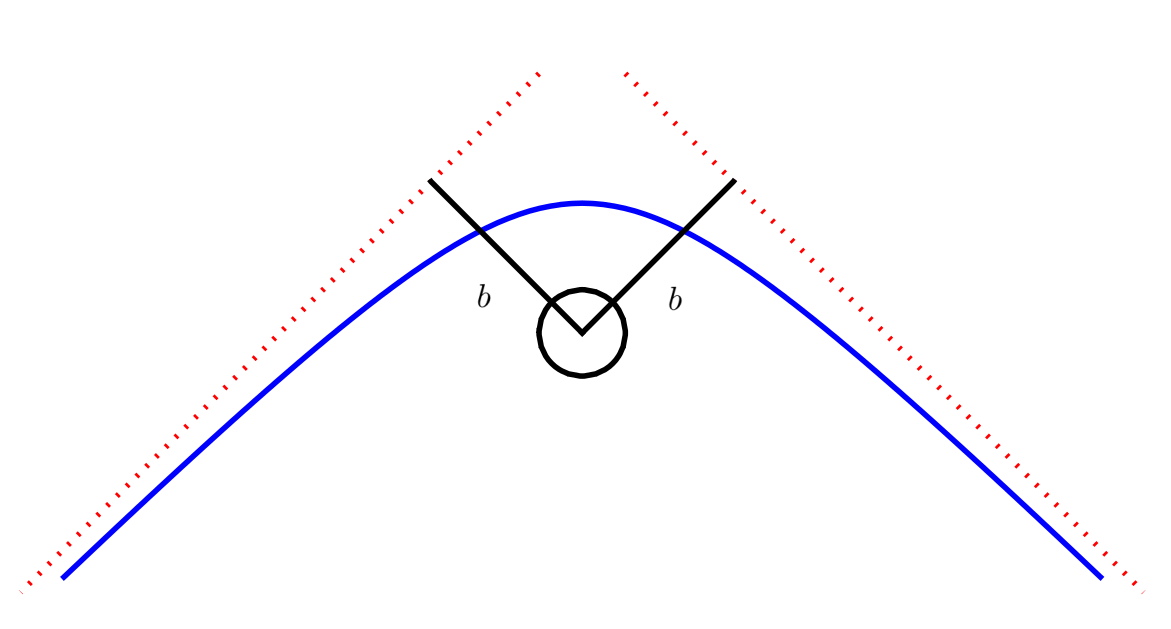}
		\caption{The impact parameter $b$ is defined as the nearest point of approach between the {\it undeviated} light ray (as seen from infinity) and the black hole. As $b$ is determined by conserved charges via $b= \frac{L}{E}$, the impact parameter for the in going and outgoing rays are equal.}
		\label{impactfig1}
	\end{figure}

	Consider an in going ray of light that starts out at infinite distance and at angular location $\phi=0$ (on the equatorial plane $\theta= \frac{\pi}{2}$) of  the 
	Schwarzschild metric 
	\begin{equation}\label{sch} 
		(ds)^2= -\left(1-\frac{r_0}{r} \right)(dt)^2 + \frac{(dr)^2}{(1-\frac{r_0}{r})}  + r^2 ((d\theta)^2 + \sin^2 \theta (d \phi)^2).
	\end{equation} 
	Consider a geodesic that starts out at $\phi=0$ when $r=\infty$ (i.e. $x=0$, see \eqref{xdef}) and then moves inwards with positive angular momentum. The equation of this trajectory is given by the well known formula (see chapters 8.4 and 8.5 of \cite{Weinberg:1972kfs})
	\begin{equation} \label{finmop}
		\phi(x) = \int_0^x \frac{dx }{\sqrt{\zeta^2 - x^2 + x^3}},
	\end{equation}
	where $\phi$ is the angle coordinate in \eqref{sch}, $x$ is a re-scaled inverse radius 
	\begin{equation}\label{xdef}
		x= \frac{r_0}{r}.
	\end{equation} 
	The constant $\zeta$ is given in terms of the 
	energy $E$ and angular momentum $L$ - or equivalently in terms of the  impact parameter $b$ (see Fig \ref{impactfig1}) by  
	\begin{equation}\label{betadef} 
		\zeta = \frac{E{r_0}}{L} = \frac{r_0}{b}.
	\end{equation} 
	
	\noindent \eqref{finmop} describes the trajectory of 
	a light ray that starts out at infinity at the angle $\phi=0$, moves inwards (and towards positive values of $\phi$) until it reaches a minimum value of the radius (i.e. a maximum value of $x=x_1$) and then moves back out to larger values of the radius (smaller values of $x$). Clearly
	the function $\phi(x)$ is double valued on its domain of definition $[0, x_1]$. We denote $\phi(x)$ on the inward journey as $\phi_{\rm on}(x)$, and on the return journey as $\phi_{\rm ret}(x)$. As the onward and return motions are `symmetrical' about the turning point $x_1$, it follows that
	\begin{equation}\label{phiretans}
		\phi_{\rm ret}(x)= 2 \phi_{\rm on}(x_1)-\phi_{\rm on}(x).
	\end{equation}
Consequently, the total angular deviation over the trip, $\Delta \phi$  is given by 
	\begin{equation}\label{deltaphi}
		\Delta \phi= \phi_{\rm ret}(0) - \phi_{\rm on}(0) = 2\phi_{\rm on}(x_1).
	\end{equation} 
	While \eqref{finmop} gives a completely explicit formula for $\phi(x)$, the integral on its RHS cannot be evaluated in terms of elementary functions for general values of $\zeta$. While Mathematica evaluates the integral in terms of EllipticF functions (see Appendix \ref{eliptic}), we have not found this result very useful.  
	While we always have the option of resorting to numerics, numerical results are often most useful when used in conjunction with all available analytic results. \footnote{In addition,  analytic results often yield powerful qualitative insights.} 
	
	The function $\phi(x)$, defined by \eqref{finmop}, is analytically tractable in the neighbourhood of atleast three special values of $\zeta$, i.e $\zeta=0$, $\zeta=\infty$ and $\zeta=\zeta_c=\frac{2}{3 \sqrt{3}}$. 
	
	The neighbourhood of $\zeta=0$ corresponds to very large impact parameters: the corresponding light trajectories are close to straight lines that pass far from the black hole. It is, thus, relatively easy to evaluate  \eqref{finmop} in a power series expansion in $\zeta$ (see e.g. \cite{Einstein:1915bz, Virbhadra:1999nm, PhysRevD.72.104006}). Such trajectories form the basis for the theory of weak lensing (see e.g. \cite{Virbhadra:1999nm} and references therein).
	
	The neighbourhood of $\zeta=\infty$ describes trajectories that are close to inward radial lines at constant $\phi$ that plunge into in the black hole: such trajectories are of limited interest to the outside observer.
	
	In this paper we focus on the neighbourhood of $\zeta=\zeta_c =\frac{2}{3 \sqrt{3}}$, i.e. on impact parameters in the neighbourhood of $b_c=\frac{3 \sqrt{3}r_0}{2}$ (recall $r_0$ is the Schwarzschild radius). At precisely this value of $b$, the integral in \eqref{finmop} is elementary, and we find the simple expression, 
	\begin{equation}\label{bgbc}
		\phi(x) =\ln \left( \frac{1+\sqrt{ x+\frac{1}{3}}}{1-\sqrt{ x+\frac{1}{3}}} \right) - \ln \left( \frac{1+\sqrt{ \frac{1}{3}}}{1-\sqrt{\frac{1}{3}}} \right).
	\end{equation}  
	Light incident on the black hole at this critical value gets trapped in a circular orbit - the so called photon ring -  located at $r= \frac{3 r_0}{2}$.
	Neighbouring trajectories on one side of  this critical ray (i.e. trajectories with  $\zeta<\zeta_c$) circle around the black hole multiple times before finally making it back out to infinity. On the other hand, trajectories with $\zeta$ just larger than $\zeta_c$ circle around the black hole multiple times before finally being sucked into the black hole singularity. The critical curve \eqref{bgbc} divides trajectories into two classes: those that escape from the black hole and those that do not. In other words, photon trajectories undergo a sort of `phase transition' at $\zeta=\zeta_c$. 
	As is usually the case, the neighbourhood of this  `phase transition' is intensely interesting \footnote{For example, it has recently been pointed out that the analogue of these trajectories around AdS Schwarzschild black holes gives rise to the intensely interesting (and surprising from the boundary viewpoint) phenomenon of Bulk Cone singularities, in the context of the AdS/CFT correspondence 
		\cite{Dodelson:2023nnr}.}
	(e.g. it is characterized by 
	a universal exponent, see below). 
	This paper is devoted to a detailed study of photon trajectories in the vicinity of 
	this phase transition.
	
	From a practical point of view, trajectories in the neighbourhood of $\zeta=\zeta_c$ are interesting because they involve very large angular deviations and so form the basis for the theory of very strong lensing: in particular for the theory of secondary images and secondary (or relativistic) Einstein rings. 
	
\begin{figure}
		\centering
		\includegraphics[width=0.6\linewidth]{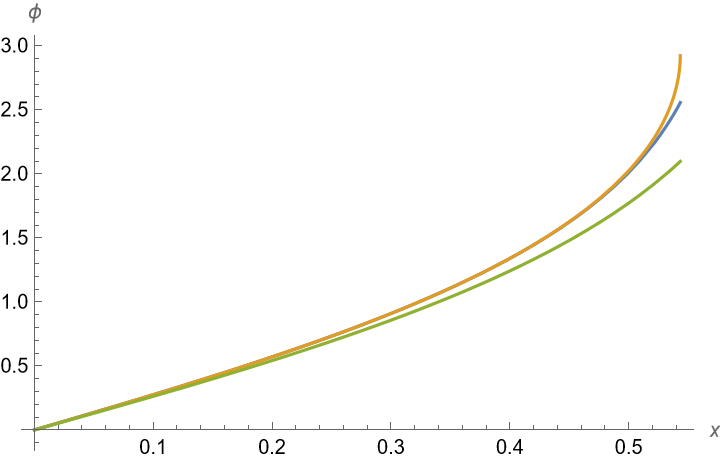}
		\caption{Graph of $\phi_{\rm on}(x)$ vs $x$ valid at small $x$, plotted at $\epsilon = 0.1$. The yellow curve is the result from numerical integration. The green curve is the previously known leading order expression for this trajectory. The blue curve is the perturbative prediction of this is paper \eqref{phionep}, valid to order $\epsilon^4$.}
		\label{phismallx}
	\end{figure}

	As trajectories at $\zeta = \zeta_c$ are so simple, (see \eqref{bgbc}), trajectories in the neighbourhood of $\zeta=\zeta_c$ are also amenable to analytic analysis. In this regime, the total angular deviation $\Delta \phi$ was first computed in \cite{Darwin} (to leading order in an expansion in $\zeta_c-\zeta$) and to higher orders in the same expansion in \cite{Iyer_2007}. At leading order, the full equation of the trajectory - i.e. a formula for the angular deviation as a function of $x$ - was largely worked out in \cite{Bozza:2007gt} \footnote{These results have been extended in many directions - and have found several applications in the papers
		\cite{Darwin,Nemiroff:1993he, Virbhadra:1999nm, Frittelli:1999yf, Eiroa:2002mk, Petters_2003, Perlick:2003vg, Bozza_2001, Bozza:2002zj,  Bozza_2004, Eiroa:2003jf, Bozza_2005, Eiroa:2004gh, 
			Bozza:2006nm,Bozza:2006sn, Eiroa:2005ag, Amore_2007, Gyulchev:2006zg, Bozza:2008ev, Chen:2009eu, 
			perlick2010gravitational, Ghosh:2010uw, Bozza:2010xqn,Bin-Nun:2010lws, Ding:2011,Aazami:2011tu, Aazami:2011tw, Wei_2012, Tsupko:2013cqa,  Sadeghi_2014, article2, Ishihara:2016sfv,Chakraborty_2017, Barlow_2017, Bozza:2017dkv, Uniyal_2018, Ono:2019hkw,PhysRevD.101.124038, PhysRevD.103.024033, Tsukamoto_2021, PhysRevD.103.124052, PhysRevD.103.044047, Aratore_2021, Bisnovatyi_Kogan_2022, Jia:2020qzt}.
	}. In this paper we build on the results of \cite{Darwin, Iyer_2007, Bozza:2007gt} to systematically work out a complete formula for photon trajectories in the neighbourhood of $\zeta=\zeta_c$ accurate to order $(\zeta_c-\zeta)^2$, improving the previous results of \cite{Bozza:2007gt} (which were valid only to order $(\zeta_c-\zeta)^0$). The substantial extent of this improvement can be seen in Fig. \ref{phismallx}. 
	
As a check of our results, we verify that the formula for $\Delta \phi$ obtained from our results agrees exactly with the formulae for the total deviation presented in \cite{Iyer_2007}. We also verify in  Appendix \ref{cani} that the deviation of our results from numerics is indeed of order $(\zeta_c-\zeta)^3$. We view these independent checks as  
	compelling evidence for the correctness of our formulae.  
	
	\begin{figure}
		\includegraphics[width=\linewidth]{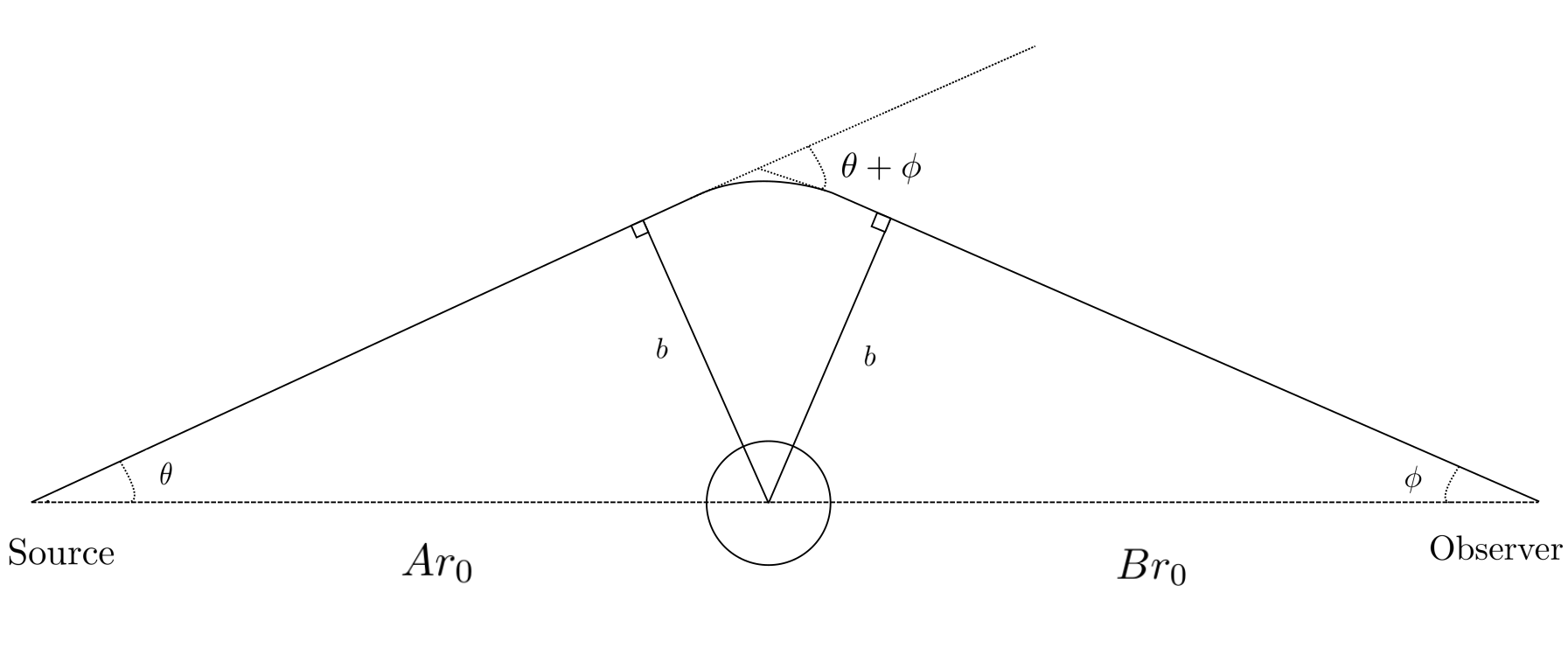}
		\caption{The passage of a ray of light from the source to the observer. In this diagram we have depicted the `direct' light ray that gives rise to the primary Einstein ring. The light rays studied in this paper actually wind $n$ times around the black hole before reaching the observer, giving rise to the $n^{th}$ relativistic Einstein Ring.}
		\label{graph3}
	\end{figure}
	With these results in hand, we study the following situation, depicted in Fig \ref{graph3}. Consider a point source, a black hole and an observer, all of which lie in a straight line. Let the black hole have Schwarzschild radius $r_0$, and let the radial distance between the source and the black hole (i.e. value of the Schwarzschild $r$ coordinate of the source) be given by $A r_0$, while the distance between the observer and black hole be $B r_0$. Let us assume that $B \gg 1$, but let $A$ take any value greater than $\frac{3}{2}$. When the observer attempts to image the source, she will observe an infinite number of Einstein rings\footnote{In Fig \ref{graph3}, we have depicted the photon trajectory that gives rise to the primary Einstein Ring. The trajectories for the secondary rings (which we have not depicted in the figure) involve the light circling the black hole.}. Let the angular radius of the $n^{th}$ ring be denoted by $\phi_n$. In the rest of this paper we use our results for the trajectories of light with $\zeta$ near $\zeta_c$ to find a formula for $\phi_n$ in the limit that $n$ is large.
	
	It turns out (and we demonstrate) that this expansion for $\phi_n$ takes takes the following form 
	\begin{equation}\label{phinexp}
		\begin{split}
			B\phi_n=&  \frac{3 \sqrt{3}}{2} +  e^{-\pi(2n+1)}P_{1,0}+e^{-2\pi(2n+1)}\Big(P_{2,0} + (2n+1) P_{2,1}\Big)+\\
			&e^{-3\pi(2n+1)}\Big(P_{3,0} + (2n+1) P_{3,1}+ (2n+1)^2 P_{3,2}\Big)+\ldots~~~,
		\end{split}
	\end{equation} 
	where $P_{i,j}$ are functions of the source distance $A$ (which we will determine below),  but are independent of $n$.
	
	In Section \ref{rings} (see \eqref{reseiring}), we present expressions for $P_{1,0}, P_{2,0}$ and $P_{2,1}$. We have also determined $P_{3,0}$, $P_{3,1}$ and $P_{3,2}$, but the final results for these quantities are lengthy, so we have explicitly presented the results for these quantities only in special limits.
	
	The expression for $P_{1,0}$ was previously presented in \cite{Bozza:2007gt}. We find that our result for $P_{1,0}$ agrees exactly with that of \cite{Bozza:2007gt} (after correcting minor typographical errors in that paper, see under \eqref{approach}). 
	
	To end this introduction we highlight (what we believe to be) the four most interesting aspects of our results. 
	\begin{itemize}
		\item When photons move in dusty environments, it is practically useful to have control over detailed photon trajectories, in order to be able to control effects like attenuation due to scattering. As the graph presented in Fig. \ref{phismallx} makes clear, the formulae of this paper represent a substantial improvement - in this respect-  over previous available analytic results. 
		\item \eqref{phinexp} presents an analytic expansion for the angular location of the $n^{th}$ Einstein ring as a function of $n$ and the radial location of the source. \footnote{The expansion parameter in this expression, $e^{-\pi(2n+1)}$ is rather small - of order $10^{-4}$ - even when $n$ takes the smallest nontrivial value, namely $n=1$. It follows that the expansion \eqref{phinexp} will be rather accurate at every nontrivial value of $n$. It also follows, however, that the resolution of the various `relativistic' Einstein rings will be practically impossible in many situations} We find the analytic form of this expression (as a function of $n$) very interesting. Recall that \eqref{phinexp} had previously only been worked out to first order. Naively extrapolating from this result, one might have been forgiven for predicting that the $q^{th}$ term in the expansion would be proportional to $e^{-q \pi (2n+1)}$. As we see from \eqref{phinexp} this is not quite the case: the `universal exponent' above is multiplied by a polynomial of $(q-1)^{th}$ order in $n$. 
		\item At the structural level, the computations presented in this paper highlight the fact that an analytic expression for the photon trajectory (near $b=b_c$ and in terms of elementary functions) can only be obtained by matching in two different regimes. One obtains simple expressions when the photon is not too near to the photon ring, and separate simple expressions when the photon is not too far from the photon ring. The two different expressions, however,  match very well in an intermediate `Goldilocks' region (neither too near nor too far from the photon ring), providing effective control everywhere. 
		\item The detailed form of trajectories in the `near photon ring' region \eqref{scond} are of particular interest, as one might expect wave corrections to geometrical optics to be particularly significant here. As far as we are aware, our paper is the first to explicitly write down the form of these trajectories even at leading order (infact we present these trajectories accurate to high orders in a perturbative expansion).
	\end{itemize}

	\section{Photon Trajectories in an expansion in $\zeta_c-\zeta$}\label{expand} 
	
	Let us suppose that 
	\begin{equation}\label{betalessmt}
		\zeta^2 = \frac{4}{27}-\frac{4}{3} \epsilon^2,
	\end{equation} 
	where $\epsilon$ is small (this equation defines $\epsilon$). We wish to evaluate the 
	function $\phi$ defined in \eqref{finmop}.
	Note that the expression under the square root in that expression is a cubic polynomial (with unit coefficient for $x^3$) and so can be 
	written as 
	\begin{equation} \label{factored}
		\zeta^2 - x^2 + x^3 = \left(x-x_1\right)\left(x-x_2\right)\left(x-x_3\right).
	\end{equation}
	It is easy to determine 
	$x_1$, $x_2$ and $x_3$ in a power series  expansion in 
	$\epsilon$.  
	We find,
	\begin{equation}\label{x123}
		\begin{split}
			& x_1 = \frac{2}{3}-\frac{2}{\sqrt{3}} \epsilon-\frac{2}{3} \epsilon ^2-\left(\frac{5}{3 \sqrt{3}}\right)\epsilon ^3-\left(\frac{16}{9}\right) \epsilon
			^4-\left(\frac{77}{12\sqrt{3}}\right)\epsilon ^5 -\left(\frac{224}{27}\right)\epsilon ^6 +{\cal O}(\epsilon^7),\\
			& x_2 = \frac{2}{3}+\frac{2}{\sqrt{3}} \epsilon-\frac{2}{3} \epsilon ^2+\left(\frac{5}{3 \sqrt{3}}\right)\epsilon ^3-\left(\frac{16}{9}\right) \epsilon
			^4+\left(\frac{77}{12\sqrt{3}}\right)\epsilon ^5 -\left(\frac{224}{27}\right)\epsilon ^6+ {\cal O}(\epsilon^7),\\
			& x_3 = -\frac{1}{3}+\frac{4}{3} \epsilon ^2+\left(\frac{32}{9}\right)\epsilon ^4+\left(\frac{448}{27}\right)\epsilon ^6 +{\cal O}(\epsilon^8).
		\end{split}
	\end{equation}
	\footnote{In particular, when $\epsilon=0$, 
		\begin{equation}\label{spvb} 
			\zeta^2 - x^2 + x^3= \frac{4}{27} - x^2 + x^3=	\left(x-\frac{2}{3}\right)^2 \left(x+\frac{1}{3}\right).
		\end{equation}
		The double root at $x=\frac{2}{3}$ is related to the fact that the photon sphere lives at $r=\frac{3r_0}{2}$.}
	In what follows we will find it  useful to work with the following linear combinations of $x_1$, $x_2$ and $x_3$, 
	\begin{equation}\label{xsumdiffnew}
		x_{\rm sum}= \frac{x_1+x_2}{2}, ~~~x_{\rm diff}= \frac{x_2-x_1}{2}, 
		~~~s^2=x_{\rm sum} -x_3.
	\end{equation} 
	Plugging \eqref{x123} into \eqref{xsumdiffnew}, we find
	\begin{equation}\label{xsumdiff}
		\begin{split}
			& x_{\rm sum} = \frac{2}{3}-\frac{2}{3} \epsilon ^2-\left(\frac{16}{9}\right) \epsilon
			^4 -\left(\frac{224}{27}\right)\epsilon ^6 +{\cal O}(\epsilon^8),\\
			& x_{\rm diff} = \frac{2}{\sqrt{3}} \epsilon +\left(\frac{5}{3 \sqrt{3}}\right)\epsilon ^3+ \left(\frac{77}{12\sqrt{3}}\right)\epsilon ^5 + {\cal O}(\epsilon^7),\\
			& s^2 = 1 -2 \epsilon ^2-\left(\frac{16}{3}\right)\epsilon ^4-\left(\frac{224}{9}\right)\epsilon ^6 +{\cal O}(\epsilon^8).
		\end{split}
	\end{equation}
	
	\subsection{Evaluation of the integral} \label{gen}
	
	The angle $\phi(x)$ is determined as a function of $x$ by 
	\eqref{finmop}. Using
	\begin{equation}\label{firstfact}
		\begin{split} 	
			&(x_1-x)(x_2-x)= \left( x_{\rm sum} -x \right)^2 - 
			{x_{\rm diff}^2},\\
			&x-x_3= s^2 -(x_{\rm sum}-x), \end{split} 
	\end{equation} 
	(see \eqref{xsumdiffnew}) it follows that the RHS of \eqref{finmop} can be rewritten as 
	\begin{equation}\label{innon}
		\begin{split} 
			s\phi(x) \equiv I(x)&=s \int_0^x dx \left(\frac{1}{\sqrt{\left( x_{\rm sum} -x \right)^2 - 
					{x_{\rm diff}^2}}}\right)\left(\frac{1}{\sqrt{s^2 - (x_{\rm sum}-x)}}\right).\\
		\end{split} 		
	\end{equation}
	Our general strategy is to evaluate the integral in \eqref{innon} after simplifying the integrand by performing power series expansions in an appropriate small parameter. We will perform two different expansions, each of which will be valid in distinct but overlapping domains of the integration variable $x$. To see how this works in more detail,  it is useful to define the variables
	\begin{equation}\label{alphabeta}\begin{split}
			&\alpha = \frac{x_{\rm diff}}{(x_{\rm sum}-x)},\\
			&\beta = \frac{(x_{\rm sum} -x)}{s^2}.\\
		\end{split}
	\end{equation} 
	Note that the combination $\alpha \beta$ is independent of $x$ as 
	\begin{equation} \label{alphabetacomb}
		\begin{split} 
			\alpha \beta &= \frac{x_{\rm diff}}{s^2}.\\
		\end{split}
	\end{equation}
	For future use we also note that when $x=0$, $\alpha$ and $\beta$ respectively take the values
	\begin{equation}\label{alphabetazero}
		\begin{split} 
			&\alpha_0\equiv \frac{x_{\rm diff}}{x_{\rm sum}},\\
			&\beta_0 \equiv \frac{x_{\rm sum}}{s^2}.\\
		\end{split}
	\end{equation}
	At the endpoint of the motion ($x=x_1=x_{\rm sum} -x_{\rm diff}$) 
	\begin{equation}\label{alphabetaend}
		\begin{split} 
			&\alpha_{\rm end}=1,\\
			&\beta_{\rm end} = \frac{x_{\rm diff}}{s^2}.\\
		\end{split}
	\end{equation} 
	It is easy to check that 
	\begin{equation}\label{thiseqab}
		(x-x_1)(x-x_2) (x-x_3)=s^2(x_{\rm sum}-x)^2(1-\alpha^2)(1-\beta),
	\end{equation}	
	so that 
	\begin{equation}\label{innonr}
		\begin{split} 
			s\phi(x)=	I(x)	&=\int_0^x \frac{dx}{(x_{\rm sum}-x)}\left(\frac{1}{\sqrt{1-\alpha^2 }}\right)\left(\frac{1}{\sqrt{1-\beta}}\right).\\
		\end{split} 		
	\end{equation}
	We will argue below that the integral $I(x)$ depends on the variable $x$, and the three constants 
	$x_{\rm sum}$, $x_{\rm diff}$ and $s$,  only through the variables $\alpha$ and $\beta$. In other words
	\begin{equation}\label{Iform}
		I=  I(\alpha, \beta).
	\end{equation} 
	For future reference, we present expressions for $\alpha$ and $\beta$, as functions of $x$ and $\epsilon$, expanded in a power series in $\epsilon$, upto the order of accuracy relevant for this paper.
	\begin{equation}\label{alphabetaep}
		\begin{split}
			&\alpha = -\frac{2 \sqrt{3}}{3 x-2}\epsilon -\frac{15 x-22 }{\sqrt{3} (3 x-2)^2}\epsilon ^3-\frac{693 x^2-1428
				x+740}{4 \sqrt{3} (3 x-2)^3} \epsilon ^5+{\cal O}\left(\epsilon ^7\right),\\
			& \beta = \left(\frac{2}{3}-x\right)-\frac{2}{3} (3 x-1) \epsilon ^2-\frac{28}{9} (3 x-1) \epsilon ^4+{\cal O}\left(\epsilon ^6\right).
		\end{split}
	\end{equation}
	
	\subsubsection{Small $\alpha$}
	
	When 
	\begin{equation}\label{fcond}
		x_{\rm sum}-x \gg x_{\rm diff},
	\end{equation} 
	it follows, from definitions, that $\alpha \ll 1$. For these values of $x$ we can evaluate $I(\alpha, \beta)$ as follows. 
	We first insert the Taylor expansion
	\begin{equation} \label{alphaexp}
		\begin{split}
			\frac{1}{\sqrt{1-\alpha^2} } =& \sum_{n=0}^{\infty} \frac{(2n)!}{(n!)^2 2^{2n}} \alpha^{2n} = 1+ \frac{1}{2} \alpha^2 + \frac{3}{8} \alpha^4  + 
			{\cal O}\left( \alpha^6\right),\\
		\end{split} 
	\end{equation}
	into \eqref{innonr} to obtain
	\begin{equation}\label{innonrao}
		\begin{split} 
			s\phi(x)=	I(x)	&=\int_0^x \frac{dx}{(x_{\rm sum}-x)}\left(\frac{1}{\sqrt{1-\beta}}\right)\left(\sum_{n=0}^{\infty} \frac{(2n)!}{(n!)^2 2^{2n}} \alpha^{2n} \right)\\
			&=\int_0^x \frac{dx}{(x_{\rm sum}-x)}\left(\frac{1}{\sqrt{1-\beta}}\right)
			\left(\sum_{n=0}^{\infty} \frac{(2n)!}{(n!)^2 2^{2n}} \frac{(\alpha\beta)^{2n}}{ \beta^{2n}} \right)\\
			&=\int_0^x \frac{dx}{(x_{\rm sum}-x)}\left(\frac{1}{\sqrt{1-\beta}}\right)
			\left( 1+ \frac{1}{2} \frac{(\alpha\beta)^2}{\beta^2} + \frac{3}{8} \frac{(\alpha \beta)^4 }{\beta^4} + 
			{\cal O}\left( \left(\frac{ \alpha \beta}{\beta} \right)^6 \right) \right).
			\\
		\end{split} 		
	\end{equation}
	Using the fact that $\alpha \beta$ is independent of $x$, we can rewrite \eqref{innonrao} as 
	\begin{equation}\label{gint}
		\begin{split}
			s\phi(x)=	I(x)=&\sum_{n=0}^{\infty} \frac{(2n)!}{(n!)^2 2^{2n}}  (\alpha \beta)^{2n} G_n(\beta)\\
			=& G_0 + \frac{\alpha^2 \beta^2}{2} G_1 + 
			\frac{3\alpha^4 \beta^4}{8} G_2 + {\cal O} \left( \alpha^6 \right),\\
		\end{split}
	\end{equation} 
	where $G_n$ are defined by 
	\begin{equation}\label{gndef} 
		G_n(\beta) =\int_0^x \frac{dx}{(x_{\rm sum}-x)}\left(\frac{1}{ \beta^{2n}\sqrt{1-\beta}}\right).	
	\end{equation} 
	Note that the RHS of \eqref{gndef} is now independent of $\alpha$. We evaluate the integral in \eqref{gndef} by changing integration variable to $\beta$. Using 
	$$ \frac{dx}{x_{\rm sum}-x} = - \frac{d\beta}{\beta},$$
	we see that \eqref{gndef} turns into 
	\begin{equation}\label{revi}
		G_n=-\int_0^{\beta} \frac{d \beta }{\beta^{2n+1}\sqrt{1-\beta}}.
	\end{equation}
	\footnote{The upper limit  $\beta$, of the integral in \eqref{revi}, is is related to the upper limit $x$ of the integral in \eqref{gndef} via the second of \eqref{alphabeta}, i.e. via $x=x_{\rm sum}-\beta s^2$ }
	The integrals $G_n(\beta)$ are easily evaluated \footnote{By moving to the variable $\sqrt{1-\beta}$.}, and we find that (at small values of $\alpha$)
	\begin{equation} \label{alphaappa} 
		I(\alpha, \beta)= f_0(\beta)+ \alpha^{2} f_1(\beta) +\alpha^4 f_2(\beta) +{\cal O}(\alpha^6),
	\end{equation}
	where 
	\begin{equation}\label{tildef}
		\begin{split}
			&f_0(\beta)=\tilde f_0(\beta)-	\tilde f_0(\beta_0),\\
			& f_1(\beta)=\tilde f_1(\beta)-	\frac{ \tilde f_1(\beta_0)\beta^2}{\beta_0^2},\\
			& f_2(\beta)=\tilde f_2(\beta)-	\frac{ \tilde f_2(\beta_0)\beta^4}{\beta_0^4},
		\end{split}
	\end{equation} 
	and 
	\begin{equation}\label{allf}
		\begin{split}
			\tilde f_0(\beta)&= \ln \left(\frac{1+\sqrt{1-\beta}}{1-\sqrt{1-\beta}} \right),\\
			\tilde f_1(\beta)&= \ln \left(\frac{1+\sqrt{1-\beta}}{1-\sqrt{1-\beta}} \right) \frac{3 \beta^2}{2^4} + ~~
			\sqrt{1-\beta}  \left( \frac{3  \beta}{8 } + \frac{1}{4} \right), \\
			\tilde f_2(\beta)&= \ln \left(\frac{1+\sqrt{1-\beta}}{1-\sqrt{1-\beta}} \right) \frac{105 \beta^4}{2^{10}}  + 
			\sqrt{1-\beta} \left( 
			\frac{105 \beta^3 }{2^{9}} 
			+ \frac{35 \beta^2}{2^{8}}  
			+\frac{7\beta}{2^{6}}  + 
			\frac{3}{2^{5}} \right). \\
		\end{split}
	\end{equation} 
	
	\subsubsection{Small $\beta$}

	When
	\begin{equation}\label{scond}
		x_{\rm sum}-x \ll  s^2
	\end{equation} 
	it follows, from definitions, that $\beta \ll 1$. For these values of $x$ it is legitimate to evaluate $I(\alpha, \beta)$ by first inserting the Taylor expansion, 
	\begin{equation}\label{thirdfactor1} 
		\begin{split} 
			\frac{1}{\sqrt{1-\beta}}=&\sum_{n=0}^{\infty} \frac{(2n)!}{(n!)^2 2^{2n}} \beta^{n} = 1 +\frac{1}{2}
			\beta + \frac{3}{8}  \beta^2 +  \frac{5}{16}
			\beta^3 + \frac{35}{128}\beta^4+ {\cal O }\left(\beta\right)^5\\
		\end{split}
	\end{equation}
	into \eqref{innonr}, to obtain, 
	\begin{equation}\label{innonrbo}
		\begin{split}
			s\phi(x) = I(x) &=\int_0^x \frac{dx}{(x_{\rm sum}-x)} \left(\frac{1}{\sqrt{1-\alpha^2}}\right)\left(\sum_{n=0}^{\infty} \frac{(2n)!}{(n!)^2 2^{2n}} \beta^{n}\right)\\ 
			&=\int_0^x \frac{dx}{(x_{\rm sum}-x)} \left(\frac{1}{\sqrt{1-\alpha^2}}\right)\left(\sum_{n=0}^{\infty} \frac{(2n)!}{(n!)^2 2^{2n}} \frac{(\alpha\beta)^{n}}{\alpha^{n}}\right)\\
			&= \int_0^x \frac{dx}{(x_{\rm sum}-x)} \left(\frac{1}{\sqrt{1-\alpha^2}}\right)\Bigg(1 +\frac{1}{2}
			\frac{(\alpha \beta)}{\alpha} + \frac{3}{8}  \frac{(\alpha \beta)^2}{\alpha^2} +  \frac{5}{16}
			\frac{(\alpha \beta)^3}{\alpha^3}\\
			&~~~~~~~~~~~~~~~~~~~~~~~~~~~~~~~~~~~~~~~~~~ + \frac{35}{128}\frac{(\alpha \beta)^4}{\alpha^4}+ {\cal O }\left(\frac{(\alpha \beta)^5}{\alpha^5}\right)\Bigg).\\
		\end{split}	
	\end{equation}
	Since $\alpha \beta$ is a constant, we can rewrite \eqref{innonrbo} as 
	\begin{equation}\label{gintbet}
		\begin{split}	
			s\phi(x)=	I(x)=&\sum_{n=0}^{\infty} \frac{(2n)!}{(n!)^2 2^{2n}}  (\alpha \beta)^{n} H_n(\alpha)\\
			=& H_0 + \frac{\alpha \beta}{2} H_1 + 
			\frac{3(\alpha \beta)^2}{8} H_2  +\frac{5 (\alpha \beta)^3}{16} H_3+\frac{35 (\alpha \beta)^4}{128} H_4+{\cal O} \left((\alpha  \beta)^5 \right),\\
		\end{split}
	\end{equation} 
	where
	\begin{equation}\label{hdef}
		H_n = \int_0^x \frac{dx}{(x_{\rm sum}-x)} \frac{1}{\alpha^n \sqrt{1-\alpha^2}}.
	\end{equation}
	Note that the RHS of \eqref{hdef} completely independent of $\beta$.  Using, 
	\begin{equation}
		\frac{dx}{(x_{\rm sum}-x)} = \frac{d \alpha}{\alpha},
	\end{equation}
	we find, 
	\begin{equation}
		H_n = \int \frac{d\alpha}{\alpha^{n+1}\sqrt{1-\alpha^2}} + c_n.
	\end{equation}
	As the `boundary condition' for the integral is at $x=0$, which lies outside the domain of the small $\beta$, $H_n$ can only be evaluated upto an unknown integration constant $c_n$. 
	In the next sub subsection we will determine the constants 
	$c_n$ by matching the expansion of this sub subsection with that of the previous sub subsection. 
	
	The integrals $H_n$ are easily evaluated \footnote{E.g. via the substitution $\alpha = {\rm sech} (\theta)$.}. We obtain 
	\begin{equation}\label{betaapp}
		I(\alpha, \beta) = g_0(\alpha)+ \beta g_1(\alpha)+ 
		\beta^2 g_2(\alpha)+ \beta^3 g_3(\alpha)+ \beta^4 g_4(\alpha) + {\cal O}(\beta^5),
	\end{equation} 
	where
	\begin{equation}\label{gall}
		\begin{split} 
			g_0(\alpha)&= -\ln \left( \frac{1 + \sqrt{1-\alpha^2}}{\alpha} \right) -c_0,\\
			g_1(\alpha)&=- \frac{\sqrt{1-\alpha^2}}{2} -c_1 \alpha,\\
			g_2(\alpha)&= -\frac{3}{16} \left(\alpha^2  \ln \left( \frac{1 + \sqrt{1-\alpha^2}}{\alpha} \right)  + \sqrt{1-\alpha^2} \right) -c_2 \alpha^2,\\
			g_3(\alpha)&= -\frac{5}{192} \left(  \sqrt{1-\alpha^2} \left(4 + 8\alpha^2\right) \right) - c_3 \alpha^3,\\
			g_4(\alpha)&=-\frac{35}{4096} \left(12 \alpha^4 \ln \left( \frac{1 + \sqrt{1-\alpha^2}}{\alpha} \right)
			+ \sqrt{1-\alpha^2}\left(8+12\alpha^2\right)\right)-c_4 \alpha^4.\\
		\end{split} 	 
	\end{equation}
	We will determine the constants $c_1$, $c_2$, $c_3$, and $c_4$  below.
	\subsubsection{Matching}

	When $\epsilon$ is small, $s \approx 1$ while $x_{\rm diff}\approx \frac{2\epsilon}{\sqrt{3}}$ (see \eqref{xsumdiff}), and so $s^2 \gg x_{\rm diff}$. It follows that there exist values of $x$ for which $x_{\rm diff} \ll x \ll s^2$. \footnote{For example \eqref{fcond} and \eqref{scond} are both parametrically obeyed for $x_{\rm sum}-x ={\cal O}(\sqrt{\epsilon})$.} At such values of $x$, $\alpha$ and $\beta$ are both small, and so the 
	expansions \eqref{alphaappa} and \eqref{betaapp} are simultaneously valid. In this sub subsection we will choose to work at such values of $x$ and so determine the (as yet unknown) integration constants 
	$c_1$, $c_2$, $c_3$, and $c_4$ (see \eqref{gall}).
	
	Taylor expanding \eqref{allf} in $\beta$
	\footnote{This Taylor expansion is justified as we here focus on values of $x$ at which $\alpha$ and $\beta$ are both small.} and inserting into \eqref{tildef} yields  
	\begin{equation}\label{powerexp} 
		\begin{split} 
			& f_0(\beta)= \left( \ln \left(\frac{4}{\beta} \right) - \tilde f_0(\beta_0)\right) - \frac{\beta}{2} -  \frac{3\beta^2}{16} 
			-  \frac{5 \beta^3}{48} - \frac{35 \beta^4}{512} +{\cal O}(\beta^5),\\
			&f_1(\beta)= \frac{1}{4}  +\frac{\beta}{4} + \frac{\beta^2}{32} 
			\left(  6\ln \left(\frac{4}{\beta} \right) -7 -\frac{32 \tilde f_1(\beta_0)}{\beta_0^2}\right)  -  \frac{5\beta^3}{32}- \frac{35 \beta^4}{512} +{\cal O}(\beta^5),\\ 
			&f_2(\beta)= \frac{3}{32}  +\frac{\beta}{16} + \frac{9\beta^2}{128} + \frac{15\beta^3}{128} + 
			\frac{\beta ^4}{4096} \left(420 \ln \left(\frac{4}{\beta }\right)-533 - \frac{4096 \tilde f_2(\beta_0)}{\beta_0^4}\right)+{\cal O}(\beta^5).\\
		\end{split} 
	\end{equation} 
	Note that while we have Taylor expanded ${\tilde f}_i(\beta)$ in $\beta$, we have not performed a similar Taylor expansion of ${\tilde f}_i(\beta_0)$. The Taylor expansion in 
	$\beta$ is justified by working at values of $x$ for which $\beta$ is small. On the other hand $\beta_0$ (see \eqref{alphabetazero}) 
	is a constant (independent of $x$) of order unity, and so a Taylor expansion in this variable is never justified.

	Similarly, Taylor expanding \eqref{gall}, to fourth order in $\alpha$,  gives 
	\begin{equation}\label{powerexpn} 
		\begin{split} 
			& g_0(\alpha)= \left( \ln \left(\frac{\alpha}{2} \right) - c_0\right)  + \frac{\alpha^2}{4} +  \frac{3\alpha^4}{32} +{\cal O}(\alpha^6),\\ 
			&g_1(\alpha)= -\frac{1}{2} -c_1 \alpha + \frac{\alpha^2}{4} + \frac{\alpha^4}{16} +{\cal O}(\alpha^6),\\
			&g_2(\alpha)= -\frac{3}{16}  -\frac{3\alpha^2}{32}\left( 2 \ln \left( \frac{2}{\alpha} \right) -1 + \frac{32 c_2}{3} \right) + 
			\frac{9\alpha^4}{128} +{\cal O}(\alpha^6),\\
			&g_3(\alpha)= -\frac{5}{48}- \frac{5 \alpha^2}{32} + \frac{15 \alpha^4}{128} -c_3 \alpha^3+{\cal O}(\alpha^6),\\ 
			&g_4(\alpha) = -\frac{35}{512} -\frac{35}{512} \alpha ^2- \frac{105}{1024} \alpha ^4 \left(\ln \left(\frac{2}{\alpha}\right) -\frac{7}{12}+ \frac{1024c_4}{105} \right) + {\cal O}(\alpha^6).\\
		\end{split} 
	\end{equation} 
	
	When $\alpha$ and $\beta$ are both small, the expression for $I$ obtained by inserting \eqref{powerexp} into 
	\eqref{alphaappa} should agree with the expression obtained by inserting \eqref{powerexpn} into \eqref{betaapp}. It is not difficult to verify that the two expressions indeed match perfectly with each other \footnote{The matching is performed in a double power series expansion in $\alpha$ and $\beta$, upto and including all terms of order ${\cal O}(\alpha^m \beta^n)$, for all $m$ and $n$ that obey the inequalities  $m \leq 4$ and $n \leq 4$.}
	if and only if the constants $c_i$ are chosen as 
	\begin{equation} \label{constmatch}
		\begin{split} 
			c_0=& 	\ln \left( \frac{x_{\rm diff}}{8 s^2} \right) + f_0(\beta_0) \\
			=&\ln \left(\frac{x_{\rm diff}\left(1+\sqrt{\frac{-x_3}{s^2}}\right)}{8s^2\left(1-\sqrt{\frac{-x_3}{s^2}}\right)}\right),\\
			c_1=&0, \\
			c_2=&\frac{5}{16} + \frac{3}{16} \ln \left( \frac{x_{\rm diff}}{8s^2}\right) + \frac{f_1(\beta_0)}{\beta_0^2}\\
			=&\frac{5}{16}+ \frac{3}{16} \ln \left( \frac{x_{\rm diff}}{8s^2}\right) + \frac{1}{\beta_0^2}\left(\sqrt{\frac{-x_3}{s^2}}\left(\frac{1}{4}  + \frac{3}{8}  \left(\frac{x_{\rm sum}}{s^2}\right)\right) + \frac{3}{2^4} \left(\frac{x_{\rm sum}}{s^2}\right)^2 \ln \left(\frac{1+\sqrt{\frac{-x_3}{s^2}}}{1-\sqrt{\frac{-x_3}{s^2}}}\right)\right),\\
			c_3=& 0,\\
			c_4=& \frac{389}{2048} + \frac{105}{1024} \ln\left(\frac{x_{\rm diff}}{8s^2}\right) +\frac{f_2(\beta_0)}{\beta_0^4} \\
			=& \frac{389}{2048} + \frac{105}{1024} \ln\left(\frac{x_{\rm diff}}{8s^2}\right) +\\
			&\frac{1}{\beta_0^4}\Bigg(\sqrt{\frac{-x_3}{s^2}} \left(\frac{3}{2^5}+ \frac{7}{2^6} \left(\frac{x_{\rm sum}}{s^2}\right) + \frac{35}{2^8}\left(\frac{x_{\rm sum}}{s^2}\right)^2 + \frac{105}{2^9}\left(\frac{x_{\rm sum}}{s^2}\right)^3\right)\\
			&~~~~~~+ 
			\frac{105}{2^{10}} \left(\frac{x_{\rm sum}}{s^2}\right)^4 \ln \left(\frac{1+\sqrt{\frac{-x_3}{s^2}}}{1-\sqrt{\frac{-x_3}{s^2}}}\right)\Bigg).\\
		\end{split}
	\end{equation}

	\section{Summary of Results for the trajectory} \label{sum}
	
	\subsection{$\phi(x)$ on the onward journey}
	
	As in the introduction we define the angular function $\phi(x)$, on the onward journey as $\phi_{\rm on}(x)$.
	
	When $\alpha \ll 1$ (i.e. when $x_{\rm sum}-x \gg x_{\rm diff}$), $s \phi_{\rm on}$ is given by the function $I(\alpha, \beta)$ listed in \eqref{alphaappa}, with $f_n(\beta)$ given by \eqref{tildef} and \eqref{allf}, and $s$ is given by the third of \eqref{xsumdiff}. 
	
	When $\beta \ll 1$ (i.e. when $x_{sum} -x \ll s^2$), 
	$s \phi_{\rm on}$ is given by \eqref{betaapp} where $g_n(\alpha)$ are listed in \eqref{gall} and $c_i$ are given by \eqref{constmatch}. Once again {\tiny }$s$ is given by the third of \eqref{xsumdiff}. 
	
	As presented in the previous two paragraphs, our final answer for
	$s\phi$ is given as a function of $\alpha$ and $\beta$. We can obtain our final result as a function of $x$ and $\epsilon$ (equivalently $x$ and $\beta$) by using the formulae  
	\eqref{alphabetaep} (which express $\alpha$ and $\beta$ as functions of $x$ and $\epsilon$). Implementing this procedure in the region $x_{\rm sum}-x\gg x_{\rm diff}$, and power expanding in $\epsilon$, we find 
	\begin{equation}\label{phionep}
		\small
		\begin{split}
			&\phi_{on}(x) = \ln \left(\frac{3+\sqrt{9 x+3}}{3-\sqrt{9 x+3}}\right)+\ln \left(2-\sqrt{3}\right)+\\
			&  \frac{\epsilon ^2}{4 (2-3 x)^2 \sqrt{3 x+1}}\Bigg[\left( -8\sqrt{3} + 90 \sqrt{3} x -90 \sqrt{3} x^2\right) + \sqrt{3x+1} \Bigg(8\sqrt{3} +20 \ln\left(2-\sqrt{3}\right) + \\
			&~~~~~~~~~~~~~~~~~20 \ln \left(\frac{3+\sqrt{9x+3}}{3-\sqrt{9x+3}}\right) - \left( 24\sqrt{3} +60 \ln \left(2-\sqrt{3}\right)+60 \ln \left(\frac{3+\sqrt{9x+3}}{3-\sqrt{9x+3}}\right)\right)x + \\
			&~~~~~~~~~~~~~~~~~\left(18\sqrt{3} +45 \ln\left(2-\sqrt{3}\right)+45\ln \left(\frac{3+\sqrt{9x+3}}{3-\sqrt{9x+3}}\right)\right)x^2\Bigg)\Bigg]+\\
			&\frac{9 \left(7-4 \sqrt{3}\right) \epsilon ^4}{16 \left(\sqrt{3}-3\right)^4 (2-3 x)^4 (3 x+1)^{\frac{3}{2}}}\Bigg[1215 x^5 \left(38 \sqrt{9 x+3}-77 \sqrt{3 x+1} \ln \left(2+\sqrt{3}\right)-154 \sqrt{3}\right)-\\
			&~~~~~~~~~~~~~~~~~~~~~~~~~~~~~~~~~~~~~~~~2835 x^4 \left(38
			\sqrt{9 x+3}-77 \sqrt{3 x+1} \ln \left(2+\sqrt{3}\right)-132 \sqrt{3}\right)+\\
			&~~~~~~~~~~~~~~~~~~~~~~~~~~~~~~~~~~~~~~~~54 x^3 \left(1520 \sqrt{9 x+3}-3080
			\sqrt{3 x+1} \ln \left(2+\sqrt{3}\right)-3773 \sqrt{3}\right)-\\
			&~~~~~~~~~~~~~~~~~~~~~~~~~~~~~~~~~~~~~~~~72 x^2 \left(190 \sqrt{9 x+3}-385 \sqrt{3 x+1} \ln
			\left(2+\sqrt{3}\right)+352 \sqrt{3}\right)-\\
			&~~~~~~~~~~~~~~~~~~~~~~~~~~~~~~~~~~~~~~~~24 x \left(380 \sqrt{9 x+3}-770 \sqrt{3 x+1} \ln
			\left(2+\sqrt{3}\right)-1749 \sqrt{3}\right)+\\
			&~~~~~~~~~~~~~~~~~~~~~~~~~~~~~~~~~~~~~~~~80 \left(38 \sqrt{3} \left(\sqrt{3 x+1}-1\right)-77 \sqrt{3 x+1} \ln
			\left(2+\sqrt{3}\right)\right)+\\
			&~~~~~~~~~~~~~~~~~~~~~~~~~~~~~~~~~~~~~~~~385 (2-3 x)^4 (3 x+1)^{3/2} \ln \left(\frac{3+\sqrt{9 x+3}}{3-\sqrt{9 x+3}}\right)\Bigg]+{\cal O} (\epsilon^6).\\
		\end{split}
	\end{equation}
	It is easy to find a similar explicit expression for $\phi_{on}(x)$ power series expanded in $\epsilon$, in the region $x_{sum} -x \ll s^2$. We do not report this expression as we will not have a use for it in this paper. 
	
	\subsection{Total angular deviation} \label{tade}
	
	The total angular deviation of the light ray, which starts and ends at infinity, is given by  
	\begin{equation}\label{dphin}
		\Delta \phi = 2 \phi_{\rm on}(x_1) = \frac{2 I(x_1)}{s}.
	\end{equation}
	When $x=x_1$, $\alpha=\alpha_{\rm end}=1$  (see \eqref{alphabetaend}). It is easy to check that the functions defined in \eqref{gall} obey $$g_0(1)=-c_0, ~~~~~g_1(1)=-c_1=0,~~~~g_2(1) = -c_2,~~~~~g_3(1) =-c_3= 0, ~~~~~g_4(1) = -c_4.$$ Using the second of \eqref{alphabetaend}, it follows that
	\begin{equation}\label{dphinn}
		\Delta \phi = -\frac{2}{r}\left( c_0 +c_2 \frac{x_{\rm diff}^2}{s^4} + c_4 \frac{x_{\rm diff}^4}{s^8} \right). 
	\end{equation}
	Plugging \eqref{constmatch} and \eqref{xsumdiff} into \eqref{dphinn}, we can explicitly find $\Delta \phi$ in terms of $\epsilon$. We find
	\begin{equation}\label{phiep}
		\begin{split}
			\Delta \phi &= \left(-2 \ln (\epsilon )-2 \ln \left(2+\sqrt{3}\right)+\ln (3)+4 \ln (2)\right)\\
			&+\frac{3}{2}\left(\frac{\left(-10 \ln
				(\epsilon )+4 \sqrt{3}-26-10 \ln \left(2+\sqrt{3}\right)+5 \ln (3)+20 \ln (2)\right)}{ \left(\sqrt{3}-3\right)^2
				\left(2+\sqrt{3}\right)}\right)\epsilon ^2\\
			&+9\left(\frac{ \left(-770 \ln (\epsilon )+380 \sqrt{3}-2157-770 \ln
				\left(2+\sqrt{3}\right)+385 \ln (3)+1540 \ln (2)\right)}{16 \left(\sqrt{3}-3\right)^4
				\left(2+\sqrt{3}\right)^2}\right)\epsilon ^4\\
			&+{\cal O}\left(\epsilon ^6\right).
		\end{split}
	\end{equation}
	Inserting 
	\begin{equation}\label{epinbeta}
		\epsilon= \sqrt{\frac{1}{9}- \frac{3 \zeta^2}{4} }= 
		\sqrt{ \frac{1}{9}- \frac{3 r_0^2}{4 b^2} },
	\end{equation} 
	(see \eqref{betalessmt}), into \eqref{phiep} yields an expression for the total angular deviation as a function of the impact parameter. \\
	\\
	The paper \cite{Iyer_2007} has also evaluated $\Delta \phi$ to order $h^{'4}$ where $h'$ is a variable, of order 
	$\epsilon$, defined in \cite{Iyer_2007}. The precise relationship between $h'$ and $\epsilon$ is recorded in the accompanying footnote.
	\footnote{In order to compare with Equation (17) of \cite{Iyer_2007}, note that $2m_{\star}^{their}$=$r_o^{our}$, $b^{their}=b^{our}$.
		$r_0^{their}= \frac{r_0^{our}}{x_1^{our}}$. $h^{their}=\frac{x_1^{our}}{2}$. $h^{'their}=1-3 h^{their}= 1-\frac{3 x_1}{2}$.(Note that $r_0^{their}$ is the $r$ coordinate of nearest approach in the orbit: the $r$ coordinate used in this paper is same as the one used in \cite{Iyer_2007}) .} We have re expressed the result of \cite{Iyer_2007} (see equation (17) of that paper) in terms of the variables employed in this paper and compared with the expression for $\Delta \phi$ presented in \eqref{phiep}. We find that the two expressions agree completely to ${\cal O} (\epsilon^4)$ (the first deviation between the two expressions is at order $\epsilon^6$). We view this perfect agreement as a highly non-trivial consistency check of the perturbation expansion presented in this paper. \footnote{In an earlier preprint version of this paper, we had reported that Equation (17) of \cite{Iyer_2007} and \eqref{phiep} agreed upto order $\epsilon^3$, but differed at order $\epsilon^4$. This incorrect claim arose from an input error we made while processing Equation (17) of \cite{Iyer_2007} on Mathematica.}
	
	\subsubsection{Total Angular Deviation at leading order }
	
	At leading order \eqref{epinbeta} becomes 
	\begin{equation}\label{epinbetafo}
		\epsilon= \sqrt{\left(\frac{3}{4} \right) } 
		\sqrt{\frac{4}{27}- \zeta^2 } 
		\approx \frac{1}{3^{\frac{1}{4}}} \sqrt{ \frac{2}{3 \sqrt{3}} -\zeta } .
	\end{equation} 
	Plugging \eqref{epinbeta} into the first line of \eqref{phiep} we find 
	\begin{equation}\label{formfordeltphi}
		\Delta \phi= -\ln \left(  \frac{2}{3 \sqrt{3}} -\zeta \right) +\ln \left( 48\sqrt{3} (7-4 \sqrt{3}) \right),
	\end{equation} 	
	in perfect agreement with, e.g., Eq (9) of \cite{Bozza:2010xqn}. 
	\footnote{The expression $-\ln (u/{\bar u}-1)$, that appears in Equation (9) of \cite{Bozza:2010xqn}, translates, in the notation of this paper to  $-\ln \left( \frac{\beta_c}{\beta}-1 \right)$. At leading order (in the approach of $\beta$ to $\beta_c$) this logarithm equals 
		$\ln \beta_c -\ln \left( \beta_c-\beta\right) $, where, to leading order,  $\beta_c=\frac{2}{3\sqrt{3}}$, to leading order. }
	
	\subsection{Angular deviation from one finite value of $r$ to another}
	
	Consider a light ray with some given value of $\zeta$, and hence of $\epsilon$. Let this light ray start out at a source, located a value of the radius $r=r_s$, and end up (on the return journey) at an observer, located at some other value $r=r_{obs}$. We define the corresponding values of $x$, i.e. $\frac{r_0}{r_s}= x_s$ and  $\frac{r_0}{r_{obs}}= x_{obs}$, 
	and  assume that $x_s-x_1 \gg \epsilon$ and also that 
	$x_{obs}-x_1 \gg \epsilon$. We assume that the ray initially moves inwards (towards smaller values of $r$), before reaching its point of nearest approach and eventually moves back out, reaching $x_{obs}$ on its way back towards infinity. Let us define the change in angle over this trajectory to be 
	$$(\Delta \phi)_{x_s, x_{obs}}.$$ \footnote{With this definition, $\Delta \phi$ defined in 
		\eqref{dphin} equals 
		\begin{equation}\label{deltaphi}
			\Delta \phi =(\Delta \phi)_{0,0}.
	\end{equation} }
	The total angular deviation for this motion is given by 
	\begin{equation}\label{dphixsxo}
		(\Delta \phi)_{x_s, x_{obs}} = \phi_{\rm ret}(x_{\rm obs}) - \phi_{\rm on}(x_{\rm s}) = 2\phi_{\rm on}(x_1) - \phi_{\rm on}(x_s) - \phi_{\rm on}(x_{\rm obs}).
	\end{equation}
	We can find an explicit expansion for this $(\Delta \phi)_{x_s, x_{obs}}$ as a power series in $\epsilon$, from \eqref{phionep} and \eqref{phiep}.
	
	At leading order, we find 
	\begin{equation}\label{phifinled}
		(\Delta \phi)_{x_s, x_{obs}} = -\ln \left( \frac{\epsilon^2}{48} \frac{\left(1+\sqrt{x_s+\frac{1}{3}}\right)^2\left(1+\sqrt{x_{\rm obs}+\frac{1}{3}}\right)^2}{\left(\frac{2}{3}-x_s\right)\left(\frac{2}{3}-x_{\rm obs}\right)}\right).
	\end{equation}
	In perfect agreement with Eq. (68) of \cite{Bozza:2007gt}.\footnote{In order to compare with \cite{Bozza:2007gt}, note that $u^{their}=b^{our}$, $(1+\epsilon^{their})^2= \frac{1}{1-9 (\epsilon^{our})^2}$. $2 M^{their} =r_0^{our}$, 
		$D_{LS}^{their}=\frac{r_0^{our}}{x_{s}^{our}}$, 
		$D_{OL}^{their}= \frac{r_0^{our}}{x_{obs}^{our}}$. }
	
	\subsection{Inverting the expression for $(\Delta \phi)_{x_s, x_{obs}}$ }
	
	In our study of Einstein rings below, we will need to find an expression for $\frac{b}{r_0}$ as a function of the exponentiated angular deviation (valid when the angular deviation is large), i.e.
	\begin{equation}\label{angexp}
		\gamma^2 = e^{-(\Delta \phi)_{x_s, x_{obs}}}.
	\end{equation} 
	This is easily accomplished via a two step procedure we briefly describe. \\
	\\
	From \eqref{dphixsxo}, it is not difficult to verify that the expression for 
	$(\Delta \phi)_{x_s, x_{obs}}$ has the following dependence on $\epsilon$, 
	\begin{equation}\label{deltphiform} 
		(\Delta \phi)_{x_s, x_{obs}}= (A_0 - 2\ln{\epsilon}) +(A_1 + B_1 \ln \epsilon ) \epsilon^2 + (A_2  + B_2 \ln \epsilon) \epsilon^4 + \ldots ~~~,
	\end{equation} 
	where $A_0, A_1, A_2, B_1, B_2$ are functions of $x_s$ and $x_{obs}$, that can be found from the relevant coefficients of \eqref{dphixsxo}.
	
	Using \eqref{angexp} it follows that 
	\begin{equation}\label{gammadef}
		\gamma^2 = e^{-(A_0 - 2\ln{\epsilon}+A_1 \epsilon^2 + B_1 \epsilon^2 \ln \epsilon + A_2 \epsilon^4 + B_2 \epsilon^4 \ln \epsilon)} = e^{-(A_0 + A_1 \epsilon^2 + A_2 \epsilon^4)} \epsilon^{(2+B_1\epsilon^2 + B_2 \epsilon^4)},
	\end{equation}
	\eqref{gammadef} can be inverted to determine $\epsilon$ as a function of $\gamma$. It is not difficult to convince oneself that the formula for $\epsilon(\gamma)$ takes the form 
	\begin{equation} \label{gammaep}
		\epsilon= K \gamma \left( 1+ \gamma^2(L_1 +L_2 \ln \gamma) + \gamma^4 (L_3 + L_4\ln\gamma + L_5 \left(\ln \gamma\right)^2)\dots \right).
	\end{equation} 
	By plugging \eqref{gammaep} into \eqref{gammadef}, expanding both sides at small $\gamma$ and equating coefficients, 
	$K$, $L_1$, $L_2$, $L_3$, $L_4$ and $L_5$ are easily determined in 
	terms of $A_0$, $A_1$, $A_2$, $B_1$ and $B_2$. We find 
	\begin{equation}\label{Epdef}
		\begin{split}
			\epsilon=& e^{\frac{A_0}{2}} \gamma  \Bigg(1+\frac{1}{2} e^{A_0}
			A_1 \gamma ^2 +\frac{1}{2} e^{A_0} B_1 \gamma ^2 \ln \left(e^{\frac{A_0}{2}}
			\gamma \right)+ \frac{1}{8} e^{2 A_0} \gamma ^4 \left(5 A_1^2+2
			A_1 B_1+4 A_2\right)+\\
			&~~~~~~~~~\frac{1}{8} e^{2 A_0} \gamma ^4 \ln
			\left(e^{\frac{A_0}{2}} \gamma \right) \left(10 A_1 B_1+2 B_1^2+4 B_2
			\right)+ \frac{5}{8} e^{2 A_0} \gamma ^4  B_1^2 \ln^2 \left(e^{\frac{A_0}{2}} \gamma\right) \Bigg).\\
		\end{split}
	\end{equation}
	Note that $\epsilon$ is proportional to $\gamma$ at leading order. It follows that $\epsilon$ is indeed small when the angular deviation is large (see \eqref{angexp}), quantitatively establishing the (intuitively obvious) fact that photon trajectories near to the critical trajectory - i.e. the subject of the perturbative expansion of previous and current section - govern large angular deviations.

	With $\epsilon$ in hand, it is easy to find $\frac{b}{r_0}$ as a function of $\gamma$. Using \eqref{betalessmt}, we find 
	\begin{equation}\label{angledef}
		\frac{b}{r_0}   = \frac{1}{\zeta} = \frac{1}{\sqrt{\frac{4}{27}-\frac{4}{3}\epsilon^2}}.
	\end{equation}
	Substituting \eqref{Epdef} into \eqref{angledef}, and power series expanding in $\gamma$, we find, 
	\begin{equation}\label{finang}
		\begin{split}
			\frac{b}{r_0} = & \frac{3 \sqrt{3}}{2}+\frac{27}{4} \sqrt{3} e^{A_0} \gamma ^2+ \frac{27}{16} \sqrt{3}
			e^{2 A_0} \gamma ^4 \left(-4 B_1 \ln \left(e^{\frac{A_0}{2}} \gamma \right)+4
			A_1+27\right)+ \\
			&\frac{27}{32} \sqrt{3} e^{3 A_0} \gamma ^6 \Bigg(-24 A_1 B_1 \ln
			\left(e^{\frac{A_0}{2}} \gamma \right)+12 B_1^2 \ln ^2\left(e^{\frac{A_0}{2}} \gamma
			\right)+4 B_1^2 \ln \left(e^{\frac{A_0}{2}} \gamma \right)-\\
			&108 B_1 \ln
			\left(e^{\frac{A_0}{2}} \gamma \right)-8 B_2 \ln \left(e^{\frac{A_0}{2}} \gamma \right)+12
			A_1^2-4 A_1 B_1+108 A_1+8 A_2+405\Bigg).
		\end{split}
	\end{equation}
	As mentioned above, the quantities $A_i$ and $B_i$ are functions of $x_s$ and $x_{obs}$, whose explicit form may be read off from \eqref{dphixsxo}. In the next section we find the explicit form of these expressions in a physically relevant context.

	\subsection{Checks Against Numerics} 
	
	In Appendix \ref{cani} we subject the the final results of our perturbative analysis (summarized in this section) to 
	a detailed quantitative comparison against the results
	from the photon trajectory obtained from a numerical integration of \eqref{finmop}. In qualitative terms we verify the excellent agreement between our perturbative results and those from numerical integration, 
	even at values of $\epsilon$ that are not very close to zero (e.g. $\epsilon=0.1$). More quantitatively, we provide numerical evidence that the error in our formula 
	\eqref{phiep} (for the total angular deviation from infinity to infinity) is of order $\epsilon^6$ as expected. In a similar manner, we  provide numerical evidence that the error in our formula for the trajectory function in the region $x_1 -x \gg \epsilon$ is of order $\epsilon^6$ (as expected), and present a similar verification for the trajectory in the region 
	$x_1 -x \ll 1$. \\
	\\
	The impressive quantitative match between our predictions 
	and numerics convinces us that the final formulae summarized in this section are accurate in all details.

	\section{The angular location of the $n^{th}$ Einstein Ring}\label{rings} 
	
	Consider a situation in which a point source (i.e. a `star'), a Schwarzschild black hole and an observer all lie in a straight line, with the black hole between the observer and the source. As in Fig. \ref{graph3}, let the radius of the event horizon of the black hole be $r_0$, let $A r_0$ represent the radial coordinate $r_s$ of the source, and let $B r_0$ represent the radial coordinate $r_{obs}$ of the observer. It follows that $x_s$ (the value of $x$ at the source) and $x_{obs}$ (the value of $x$ at the observer) are given by 
	\begin{equation}\label{xsetc} 
		x_s= \frac{r_0}{A r_0}=\frac{1}{A}, ~~~~x_{obs}= \frac{r_0}{Br_0}=\frac{1}{B}.
	\end{equation} 
	In this section we assume that $B \gg 1$, but allow $A$ to take any value (subject to the requirement that $A$ lie outside the photon ring, i.e. $A> \frac{3}{2}$).\\
	\\
	Consider a ray of light emitted from the source,  circles the black hole $n$ times and finally reaches the observer. The total change in angle on the trajectory of this ray - between its emission at $x=x_s$ and its absorption at $x=x_{obs}$  - is $\pi(2n+1)$. Let the impact parameter of this light ray equal $b_n$. It follows from elementary geometry that this ray (along with its counterparts that are obtained by rotating around the axis of symmetry of this problem) would appear, to the observer, as a ring of angular radius 
	\begin{equation}\label{phin}
		\phi_n= \sin^{-1}\left(\frac{b_n}{r_0 B}\right) \approx \frac{b_n}{r_0 B}.
	\end{equation} 
	Of course $b_n$ is not a free parameter. It is determined by the parameters $x_s$, $x_{obs}$ and $n$. $n$ can only be large when $b_n$ is just greater than the critical impact parameter. In this situation, the analysis of the previous section applies, and $\frac{b_n}{r_0}$ is given by \eqref{finang} with $\gamma^2=e^{-\pi(2n+1)}$. To make 
	\eqref{finang} concrete, we substitute in the expressions for  $A_0, A_1...$ (from \eqref{deltphiform} and \eqref{dphixsxo} - after setting $x_{\rm obs} = \frac{1}{B}$ to zero. Recall that  $x_s=\frac{1}{A}$ is some finite number. We find 
	\begin{equation} \label{reseiring}
		\small
		\begin{split}
			B\phi_n = &	\frac{3 \sqrt{3}}{2}\\
			&-\gamma^2 \frac{972 \left(\sqrt{3}-2\right)  \left(\sqrt{3}-\sqrt{3
					x_s+1}\right)}{\sqrt{3} \sqrt{3 x_s+1}+3}+\gamma^4\frac{3888 \sqrt{3} \left(4 \sqrt{3}-7\right)  \left(\sqrt{3} \sqrt{3
					x_s+1}-3\right)^2}{(3 x_s-2)^2 \sqrt{3 x_s+1} \left(\sqrt{3} \sqrt{3 x_s+1}+3\right)^2}\times\\
			& \Bigg[ \sqrt{3 x_s+1} \Bigg(-\left(90 x_s^2-120 x_s+40\right) \ln \left(\frac{\gamma }{\sqrt{\frac{\sqrt{9
							x_s+3}+3}{\left(\sqrt{3}-2\right) \left(\sqrt{9 x_s+3}-3\right)}}}\right)\\
			&+\left(9+18 \sqrt{3}\right) x_s^2-\left(45
			x_s^2-60 x_s+20\right) \ln \left(-\frac{\left(2+\sqrt{3}\right) \left(\sqrt{9 x_s+3}+3\right)}{\sqrt{9
					x_s+3}-3}\right)\\
			&-\left(12+24 \sqrt{3}\right) x_s+8 \sqrt{3}+4\Bigg)+90 \sqrt{3} x_s^2-90 \sqrt{3} x_s+8 \sqrt{3}\Bigg]\\
			&+ \gamma^6 \left( \phi_n^{(6)} (x_s)+ \phi_n^{\ln (6)}(x_s) \ln \gamma +\phi_n^{\ln^2 (6)}(x_s) \ln^2 \gamma\right) \\
			&+{\cal O}(\gamma^8).\\
		\end{split}
	\end{equation}
\footnote{Note that the square bracket listed in the third, fourth and fifth lines of \eqref{reseiring} multiplies the second expression - proportional to $\gamma^4$ in the second line of the same equation.} We have also found explicit expressions for the functions 
	$\phi_n^{(6)} (x_s)$, $\phi_n^{\ln, (6)}(x_s)$, and $\phi_n^{\ln^2 (6)}(x_s)$ but have not reported our results for these quantities in general, as the expressions are rather lengthy. The expressions for these functions simplify in the limit 
	$x_s\rightarrow 0$ (i.e. when the source is very far away). In this special limit we find  
	\begin{equation}\label{repphin}
		\small
		\begin{split} 
			&\phi_n^{(6)} (0)=-11664 (-2340 + 1351 \sqrt{3}) \left(-331 + 84 \sqrt{3} + 
			150 \left(4 (\cos^{-1}(2))^2 + \left(\ln({7 - 4 \sqrt{3}})\right)^2\right)\right)\\
			&~~~~~~~~~\approx 1387.03,\\
			&\phi_n^{\ln (6)}(0)=349920 (-95268 + 55003 \sqrt{3})\approx -3300.2,\\
			&\phi_n^{\ln^2 (6)}(0)=6998400 (-2340 + 1351\sqrt{3}) \approx 4486.15~.\\
		\end{split} 
	\end{equation} 		
	In obtaining the expressions presented above, we have used the values of $A_0, A_1, A_2, B_1, B_2$ (see \eqref{deltphiform}) obtained from \eqref{dphixsxo} obtained 
	by using the approximation \eqref{alphaappa} (rather than \eqref{betaapp}) for $\phi_{on}(x_s)$. 
	Recall that \eqref{alphaappa} is valid provided 
	$x_1-x_s \gg \epsilon$. In the current situation, $\epsilon \sim \gamma \sim e^{-\frac{\pi(2n+1)}{2}}$. In other words this approximation we have used is valid provided the separation between the source and the photon sphere stays fixed (at any value, no matter how small)
	as $n$ is taken to infinity.\\
	\\
	Although likely physically irrelevant, one could ask the following mathematical question. What is the angular location of the Einstein ring as the location of the source is moved nearer and nearer to the photon sphere (as $n$ is taken to $\infty$), in a manner so that this distance scales like $e^{-\frac{\pi(2n+1)}{2}}$? In order to address this (perhaps physically artificial) question, one should use the expression \eqref{betaapp} (rather than \eqref{alphaappa}) to accurately approximate $\phi_{on}(x_s)$ in \eqref{dphixsxo}. For completeness we have also implemented this procedure to leading order. Our complete leading order result (valid for all values of $x_s$) is 
	\begin{equation}\label{approach}
		B \phi_n -\frac{3 \sqrt{3}}{2}	= 
		\begin{cases} 
			324 \sqrt{3}  e^{-(2n+1) \pi} \frac{(2-\sqrt{3})(1-\sqrt{\frac{1}{A}+\frac{1}{3}})}{(1+\sqrt{\frac{1}{A}+\frac{1}{3}})}, & \left(\frac{1}{A}-\frac{2}{3} \right) \gg e^{-\frac{\pi(2n+1)}{2}}\\
			81 \sqrt{3} e^{-(2n+1)\pi} (2-\sqrt{3})
			\left(\frac{2}{3}-\frac{1}{A} \right)
			& \left(\frac{1}{A}-\frac{2}{3} \right) \ll  1.\\
		\end{cases}
	\end{equation}
	(The first line of \eqref{approach} is a slight rewriting of the second line of \eqref{reseiring}).

	A leading order formula for the angular extent of Einstein rings, due to a source whose radial location obeys $\left(\frac{1}{A}-\frac{2}{3} \right) \gg x_{\rm diff}$ has
	previously been presented in \cite{Bozza:2007gt} (see equations (70) and (71) of that paper). The formulae (70) and (71) written in \cite{Bozza:2007gt} clearly suffer from some minor typographical errors (the formulae, as reported in \cite{Bozza:2007gt}, are dimensionally inconsistent). Starting with equations (66), (67) and (68) of that paper, we have re derived equations (70) and (71) of \cite{Bozza:2007gt}. We 
	present the corrected version of these equations in the footnote to this sentence. \footnote{Equation (71) of \cite{Bozza:2007gt} is dimensionally inconsistent as reported. This is because $D_{LS}$ has the dimensions of length, so the combinations $(2 D_{LS}-3)$ and $3+D_{LS}$, that appear in (71) of \cite{Bozza:2007gt} are  dimensionally in homogeneous. Upon performing the re derivation mentioned in the main text, we find that 
		equation (71) is correct once we replace $(2 D_{LS}-3)$ with  $(2 D_{LS}-3 (2M))$ and also replace $(3+D_{LS})$ with $\left( 3 (2M)+D_{LS} \right) $. This replacement renders the equations dimensionally consistent. As mentioned in the main text, the corrected version of Eq (71) of \cite{Bozza:2007gt} agrees perfectly with the first of \eqref{approach}. Note that $2M$, in the notation of \cite{Bozza:2007gt}, equals $r_0$ in the notation of this paper.} The first of \eqref{approach} agrees precisely with the corrected version of the formulae presented in the previous footnote.
	
	In the limit $A \to \infty$ we have 
	\begin{equation}\label{approachass}
		B \phi_n -\frac{3 \sqrt{3}}{2}	= 
		324 \sqrt{3} e^{-(2n+1) \pi} (2-\sqrt{3})^2,
	\end{equation} 
	in agreement with the the result quoted in e.g. \cite{Bisnovatyi_Kogan_2008}.

	\section{Summary and Discussion} \label{back}

	In this paper we have used the method of matched asymptotic expansions to find a detailed formula for the light like trajectories in a Schwarzschild black hole background in the neighbourhood of the critical impact parameter. We have used our results to correct minor typographical errors in the leading large $n$ formula for the radial extent of the $n^{th}$ Einstein ring \cite{Bozza:2007gt} (due to a source that is not necessarily far away from the black hole), and to improve this formula by working out higher order corrections. 
	
	We have presented formulae for the photon trajectories in the far region \eqref{alphaappa}, at orders $(b-b_c)^0$, $(b-b_c)^1$ and $(b-b_c)^2$ ($b$ is the impact parameter, and $b_c$ is its critical value).  The result at order $(b-b_c)^0$ had effectively been previously worked out in \cite{Bozza:2007gt}. Our results at this order agree with those of \cite{Bozza:2007gt}. 
	To the best of our knowledge, our results at orders $(b-b_c)^1$ and $(b-b_c)^2$ are new.
	
	As far as we are aware, the formulae for photon trajectories in the near region \eqref{scond} have never been completely\footnote{The paper \cite{Jia:2020qzt} evaluates the integral in the near region, but only upto integration constants. This paper does not perform the matching with far region results needed to determine these constants.} worked out before, even at leading order. We thus believe that both the leading order - as well as the perturbative corrections - of our expansion 
	\eqref{betaapp} are new. 
	
	As mentioned above, we have used our results for photon trajectories in the neighbourhood of the critical impact parameter to derive a formula for the angular radius $\phi_n$ of the $n^{th}$ Einstein ring resulting from a source that lies behind a black hole, exactly on the axis connecting the observer to the black hole. We have demonstrated that  the analytic dependence of $\phi_n$ on $n$ takes the interesting form given in  \eqref{phinexp}, and have found the first few coefficients in this expansion.

	When the source deviates off the axis connecting the observer to the black hole, the infinite series of Einstein rings degenerate into an infinite series of pairs of dot like images (that lie in the plane containing the observer, black hole and source - see e.g. \cite{Virbhadra:1999nm}) . It would be straightforward to use the results of this paper to compute the angular locations of the $n^{th}$ pair of images (in the large $n$ limit), accurate to third order in the small parameter $e^{-\pi (2n+1)}$. We leave this extension to the interested reader. 
	
	In this paper we have worked in the limit of geometrical optics. As the fate of geodesics in the neighbourhood of the critical trajectory depends sensitively on its impact parameter, we expect wave effects to be important for these trajectories.  It would be interesting to find wave analogues of the formulae derived in this paper. 
	
	\section{Acknowledgments} 
	
	I have benefited from very useful discussions with S. Kumar, S. Minwalla and  S. Raju. I would also like to thank G. Horowitz, G. Mandal, S. Minwalla and S. Raju for very useful comments on a draft of the manuscript. I have received no financial support towards the work contained in this paper.

	\appendix

	\section{The integral in terms of EllipticF functions}\label{eliptic}
	
	When requested to evaluate the integral \eqref{finmop} in the form 
	\begin{equation} \label{finmopxoxt}
		\int\frac{dx }{\sqrt{(x-x_1)(x-x_2)(x-x_3)}}, 
	\end{equation}
	Mathematica outputs the result 
	\begin{equation}\label{elipans}
		-\frac{2}{\sqrt{(x_2-x_1)(x-x_1)} } {\rm EllipticF}\left[ \sin^{-1} \left( \frac{\sqrt{x_2-x_1}}{\sqrt{x-x_1}} \right) , \frac{x_1-x_3}{x_1-x_2} \right].
	\end{equation} 
	We were, however, unable to induce Mathematica to Taylor expand this result around $x_1=x_2$.\footnote{In the special case $x=x_1$, Iyer and Petters \cite{Iyer_2007} were able to transform this EllipticF function into a form amenable to Taylor expansion.} 
	
	The expansions presented in Section \ref{expand} may thus be thought of as a first principles derivation of the `Taylor expansions' of the EllipticF function presented above. This expansion takes an interesting form.
	
	Our final answer for this `Taylor Expansion' (valid upto arbitrary order) is presented in \eqref{gint}, \eqref{gndef}, \eqref{gintbet} and \eqref{hdef}. The sense in which these expressions constitute a `Taylor expansion' in $\epsilon$ is the following. 
	\eqref{gint}, \eqref{gndef} yield an all orders expansion of the function valid whenever $\alpha$ is small. On the other hand, \eqref{gintbet} and \eqref{hdef} yield a second all orders expansion (initially with undetermined integration constants) valid when $\beta$ is small.
	The role that small $\epsilon$ plays is the following. The two independent expansions described above have a matching domain of  validity only when $\epsilon$ is small. In particular, the integration constants in the small $\beta$ expansion can only be determined in a power series expansion in $(\alpha \beta)^2$, which is of order $\epsilon^2$. 
	
	The elaborate structural nature of this asymptotic expansion likely explains why Mathematica finds it difficult to simply `Taylor Expand' \eqref{elipans} in $\epsilon$.

	\section{Comparison Against Numerical Integration} \label{cani}
	
	In this Appendix we compare the final results of Section \ref{sum} against results obtained for the same quantities found via numerical integration performed on Mathematica. 
	
	\subsection{The Total Angular Deviation} \label{tad}
	
\begin{figure}
		\centering
		\includegraphics[width=\linewidth]{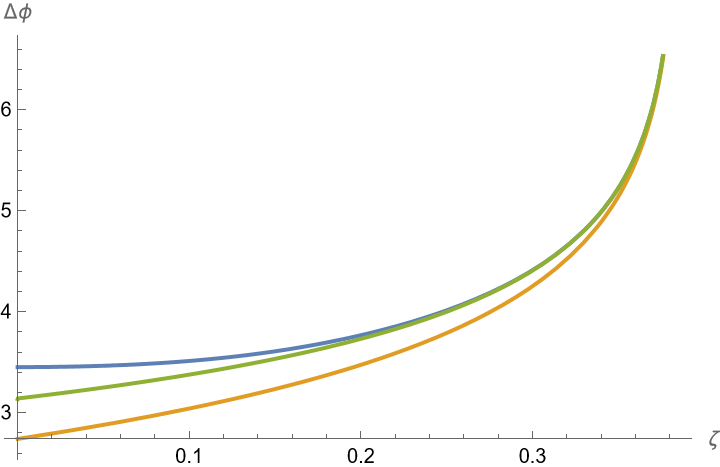}
		\caption{Plots of the total angular deviation of a light ray from infinity to infinity, as a function of $\zeta$. The graph in blue plots the final perturbative result of this paper, i.e. \eqref{phiep}. The graph in orange plots the leading order perturbative result, \eqref{formfordeltphi}. The graph in green is the result obtained from numerical integration.}
		\label{totdevcom}
	\end{figure}

	We have numerically evaluated the total angular deviation, 
	$\Delta \phi$, as a function of $\zeta$ as follows. We first used Mathematica to numerically solve the equation $\zeta^2-x^2 +x^3=0$ and so evaluated $x_1$ as a function of $\zeta$. We then numerically evaluated the integral \eqref{finmop} with upper limit $x_1$\footnote{In practice, Mathematica evaluates $x_1$ with an error $\delta \lesssim 10^{-15}$. When the error is positive, instructing Mathematica to evaluate the definite integral all the way upto $x_1$ yields a answer, as Mathematica is instructed to evaluate the square root in the integrand where it is negative. In order to avoid this issue, we instructed Mathematica to use evaluate the integral with the upper limit replaced by $x_1 + \frac{\delta}{\frac{d}{dx}\left(x^3-x^2+\zeta^2\right)\Big|_{x=x_1}}$, where $x_1$ is the root outputted by Mathematica. This fixes the problem. (Note the derivative in the denominator is negative).}. We obtained the total deviation by doubling this integral.\\
	\\
	In Fig \ref{totdevcom} we present plots of $\Delta \phi$ computed in this paper (\eqref{phiep} with the replacement 
	\eqref{epinbeta}), the leading order approximation to $\Delta \phi$ (i.e. \eqref{formfordeltphi}: equivalently the result presented in equation (9) of \cite{Bozza:2010xqn}) and the numerical result for $\Delta \phi$ (see the previous paragraph). Fig. \ref{totdevcom} illustrates that the fourth order result obtained in this paper is a substantial improvement over the leading perturbative result.
	
\begin{figure}
	\centering
	\begin{subfigure}[b]{0.4\linewidth} 
		\includegraphics[width=\linewidth]{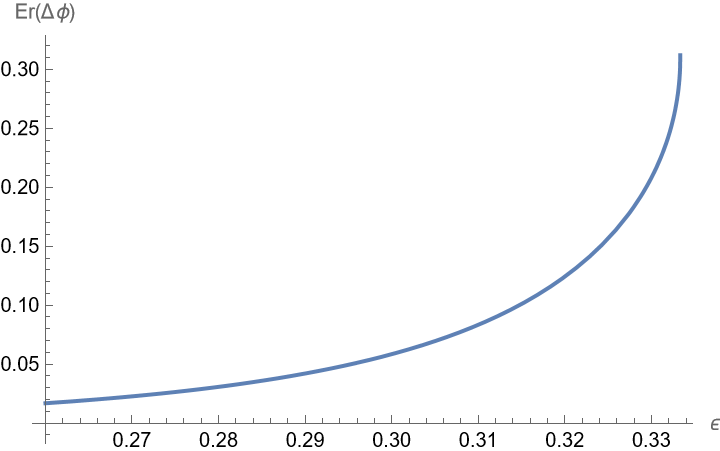}
		\caption{ Graph of error (Analytic - Numeric), of the total angular deviation vs $\epsilon$}
		\label{error}
	\end{subfigure}
	\hfill
	\begin{subfigure}[b]{0.4\linewidth}
		\includegraphics[width=\linewidth]{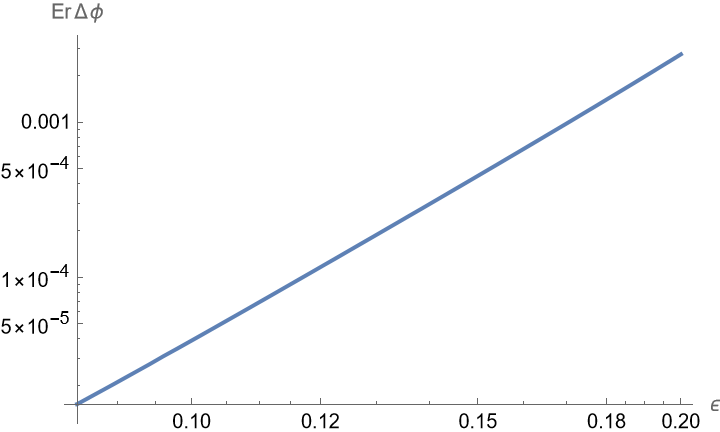}
		\caption{Log Log plot for the error in the total angular deviation vs $\epsilon$, in the range ($\epsilon=0.085$ to $\epsilon = 0.2$)}
		\label{logerror}
	\end{subfigure}
	\hfill
	\caption{ }
	\label{}
\end{figure}

	 In Fig. \ref{error} we present a plot of the error $Er(\Delta \phi)$ - defined as the total deviation as computed by \eqref{phiep} minus the same quantity obtained by numerical integration - versus $\epsilon$. We see from the figure that the error is extremely small at small $\epsilon$. 

	In Fig. \ref{logerror} we present a Log Log plot of the same graph, and find 
	an (approximately) straight line with 
	slope approximately equal to 6 \footnote{Using the values at $\epsilon=0.10$ and $\epsilon=0.13$ we obtained a slope of about 5.99.} in agreement with the expectation that the error in \eqref{phiep} is of order $\epsilon^6$.
	
	\subsection{The trajectory for $x-x_1 \gg x_{\rm diff}$}
	
	In Fig \ref{phismallx}, in the introduction, we have already presented a plot of our `small $x$' perturbative prediction for the trajectory function \eqref{alphaappa}, \eqref{gall}, \eqref{constmatch}, versus the numerical result for the same trajectory (obtained simply by numerically integrating \eqref{finmop}) at $\epsilon =0.1$. Note that the agreement is excellent: the two graphs are indistinguishable from each other for $x<0.45$. 
	
\begin{figure}
		\centering
		\begin{subfigure}[b]{0.4\linewidth} 
			\includegraphics[width=\linewidth]{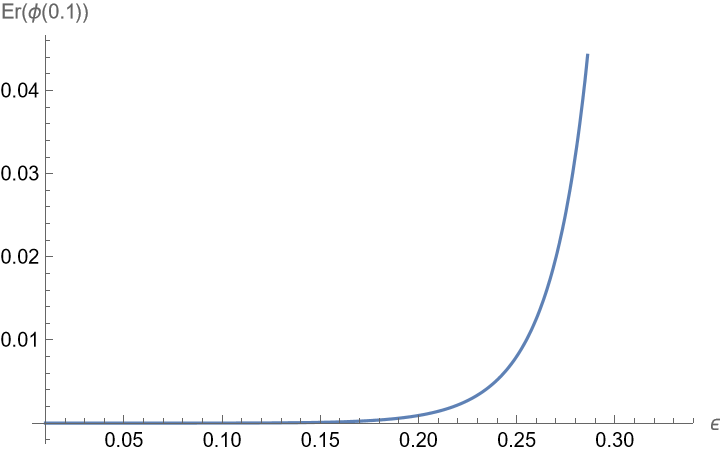}
			\caption{Graph of the error (Numeric-Analytic) of $\phi_{\rm on}(x)$ vs $\epsilon$ evaluated at $x=0.1$. }
			\label{error1}
		\end{subfigure}
		\hfill
		\begin{subfigure}[b]{0.4\linewidth}
			\includegraphics[width=\linewidth]{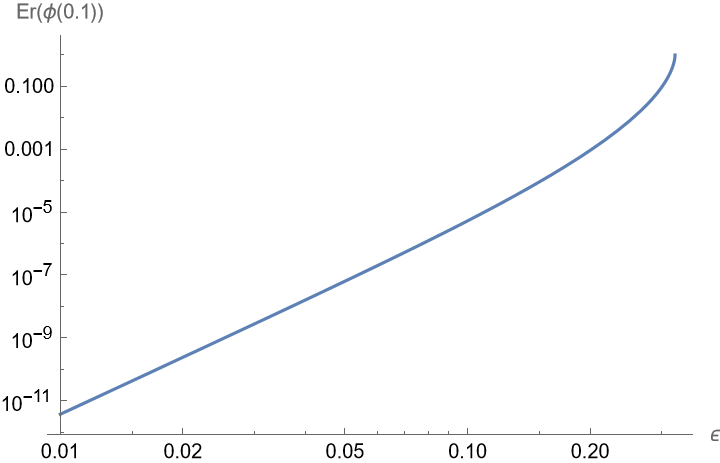}
			\caption{Log Log Graph for the error in $\phi_{\rm on}(x)$ vs $\epsilon$ evaluated at $x=0.1$. }
			\label{logerror1}
		\end{subfigure}
		\hfill
		\caption{ }
		\label{}
	\end{figure}

	In Fig. \ref{error1} we have plotted the error in $\phi_{\rm on}(x)$ at a particular value of $x$ - we chose $x=0.1$ - as a function of $\epsilon$. The error, $Er(\phi_{\rm on}(0.1))$ is defined to be the numerically obtained value minus the perturbative value for $\phi_{\rm on}(x)$ at $x=0.1$, as a function of $\epsilon$. Note that the error is extremely small at small $\epsilon$. 
	In order to obtain a quantitative estimate for the error, 
	in Fig \ref{logerror1} we have presented a log log plot for the same quantity (error versus $\epsilon$). Note that the graph is a straight line at small $\epsilon$. We estimated the slope of this graph (using the points at  $\epsilon=0.01$ and $0.03$) and found slope $=6.026 \approx 6$, in agreement with the expectation that the first correction to this result is at $\epsilon^6$. 

	\subsection{Trajectory for $x_1-x \ll 1$}
	
\begin{figure}
	\centering
	\begin{subfigure}[b]{0.5\linewidth} 
		\includegraphics[width=\linewidth]{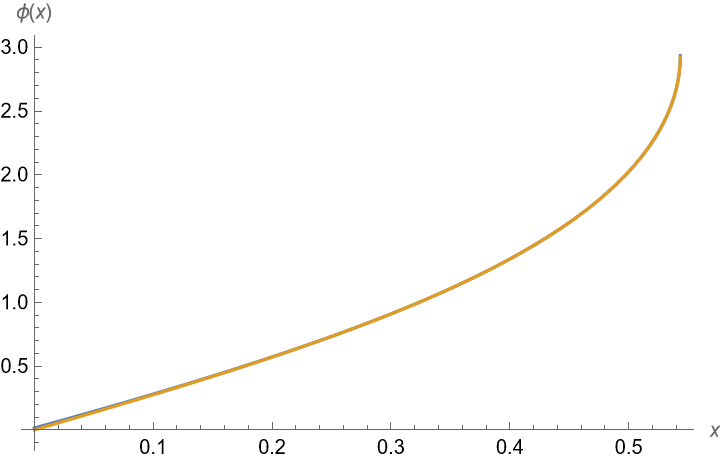}
		\caption{Graph of $\phi_{\rm on}(x)$ vs $x$ plotted at $\epsilon = 0.1$.  The yellow curve is the result from numerical integration, while the blue curve is the perturbative prediction of this is paper. Note that the two curves are virtually indistinguishable.}
		\label{philargex}
	\end{subfigure}
	\newline
	\hfill
	\begin{subfigure}[b]{0.45\linewidth}
		\includegraphics[width=\linewidth]{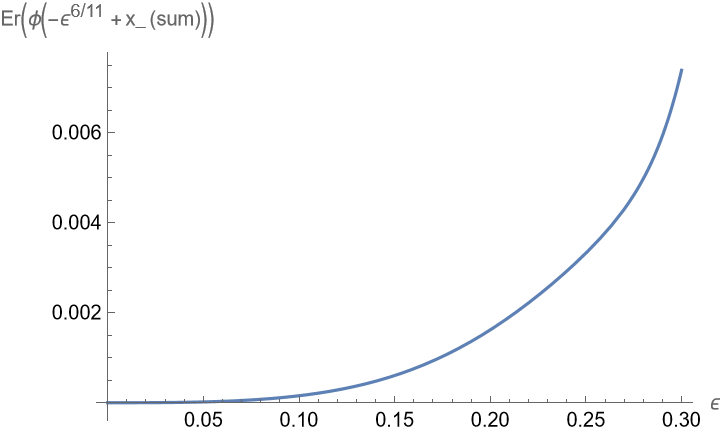}
		\caption{Graph of the error (Analytic-Numeric) of the error in $\phi_{\rm on}(x_{\rm sum}-\epsilon^{\frac{6}{11}})$ vs $\epsilon$}
		\label{error3}
	\end{subfigure}
	\hfill
	\begin{subfigure}[b]{0.45\linewidth}
		\includegraphics[width=\linewidth]{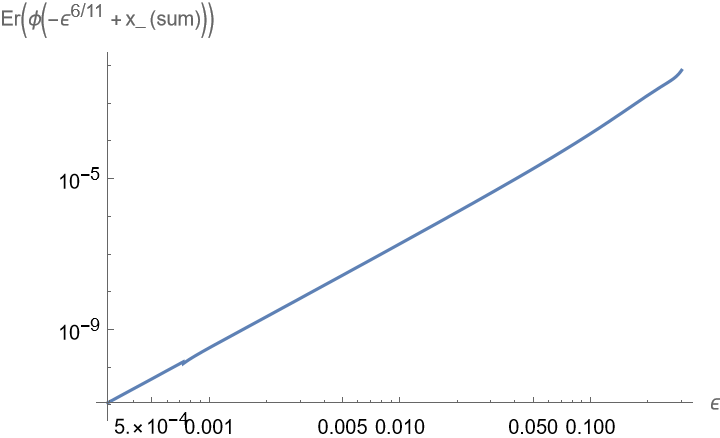}
		\caption{Log Log Graph of the error vs $\epsilon$}
			\label{logerror3}
		\end{subfigure}
		\caption{}
		\label{}
	\end{figure}

	In the region $x_1-x \ll 1$ we have computed the trajectory 
	in a power series expansion in $\beta$. In Fig \ref{philargex} we present a plot comparing our perturbative 
	prediction for the angle $\phi_{\rm on}(x$ (\eqref{betaapp}, \eqref{gall}, and \eqref{constmatch}) versus the numerical prediction for the same quantity at $\epsilon=0.1$. The agreement is excellent, even down to $x=0$. \\
	\\
	As $\beta$ does not go to zero in the limit $\epsilon \to 0$, the error in our perturbative trajectory (as computed by subtracting our perturbative results from those of numerical integration) is not expected to go to zero if $\epsilon$ is taken to zero at a fixed value of $x$. \\
	\\
	Our perturbative result in this region has two sources of errors. First we have errors of order $\alpha^6$ (because we have only matched our answer to order $\alpha^4$). Second, we have errors of order $\beta^5$ (because we have evaluated our perturbative expansion only to order $\beta^4$). If we choose to set 
	\begin{equation}\label{xsumx}
		x_{\rm sum}-x= \epsilon^q
	\end{equation} 
	then we find that $\alpha^6 \sim \epsilon^{6(1-q)}$ while 
	$\beta^5\sim \epsilon^{5q}$. We get the best accuracy when these two errors are of the same order. This occurs when 
	\begin{equation}\label{valq} 
		q=\frac{6}{11}
	\end{equation} 	
	With this choice of $q$, both our errors are expected to be of order 
	\begin{equation}\label{epthirtele}
		\epsilon^{\frac{30}{11}} =\epsilon^{2.727272...}
	\end{equation} 
	We now proceed to quantitatively verify that our perturbative prediction for the error takes this form. 

	In Fig \ref{error3} we present a plot of the error in 
	$ \phi_{\rm on}(x_{\rm sum} -\epsilon^{\frac{6}{11}})$
	versus $\epsilon$. The error is defined as our analytic perturbative prediction \eqref{betaapp} minus the same quantity evaluated from numerical integration. As expected, the error vanishes rapidly at small $\epsilon$. In order to make this statement quantitative, we present a log log plot of the same curve 
	in Fig \ref{logerror3}. As expected, the curve is a straight
	line, whose slope (evaluated using the points $\epsilon=0.001$ and $\epsilon=0.005$) is approximately 
	2.75, within a percent of our expectation of slope equal 
	to 2.727272. 
	\newpage
	\bibliography{sn-bibliography}
\end{document}